\documentclass[preprint,aps,superscriptaddress,nofootinbib]{revtex4}
\usepackage{amsmath}
\usepackage{graphicx}
\usepackage{color}
\usepackage{slashed}

\newcommand{\vsq}{\langle v^2 \rangle}
\newcommand{\vev}[1]{\left< #1 \right> }
\newcommand{\nn}{\nonumber }

\begin{document}

\title{Extremal noncommutative black holes as dark matter furnaces}
\author{Shoichi Kawamoto}\thanks{%
E-mail: kawamoto@cycu.edu.tw}
\affiliation{Department of Physics and Center for High Energy Physics, Chung Yuan Christian University, Chung Li, Taoyuan 320, Taiwan}
\author{Chun-Yu Wei}\thanks{%
E-mail: weijuneyu@gmail.com}
\affiliation{Department of Physics and Center for High Energy Physics, Chung Yuan Christian University, Chung Li, Taoyuan 320, Taiwan}
\author{Wen-Yu Wen}\thanks{%
E-mail: wenw@cycu.edu.tw}
\affiliation{Department of Physics and Center for High Energy Physics, Chung Yuan Christian University, Chung Li, Taoyuan 320, Taiwan}
\affiliation{Leung Center for Cosmology and Particle Astrophysics\\
National Taiwan University, Taipei 106, Taiwan}

\begin{abstract}
In this letter, we consider dark matter annihilation in the gravitational field of noncommutative black holes.  At final stage of evaporation, we hypothesize the existence of a thermal equilibrium state composed of a burning black hole relics fueled by dark matter accretion.
\end{abstract}

\maketitle



\section{Introduction}
The Dark Matter is estimated to contribute about $26.8\%$ content of
our universe and abundant in galaxies as halos.  The Dark Matter
Particles are believed to accrete and affect various annihilation
channels significantly on the supermassive black hole at the center of
each galaxy\cite{Baushev:2008yz}.   There are  estimated certain
amount of primordial black holes (PBHs)
 remained since their creation in the
very early universe.  Those tiny relics may possess mass range from
$10^{14}$ to $10^{23}$kg and their possible role as a Dark Matter
candidate has been widely discussed\footnote{Readers are directed to
  \cite{Green:2014faa, Gelmini:2015zpa} for a review.}.  However,
regardless any chance to be dark matter candidate, the fate of those
tiny black holes remains unclear due to the lack of full knowledge of
Planck scale physics in curved spacetime.  The conventional thermal
description of black holes seems doomed to failure due to the
unphysical infinite Hawking temperature for infinitesimal mass.
Several proposals to modify the fundamental properties of spacetime,
such as General Uncertainty Principle (GUP) and noncommutative
geometry (NG), is to introduce a new scale when the black hole could
stop evaporating about Planck size.  In particular, a NG or GUP
inspired Schwarzschild black hole would only reach up to sub-Planck
temperature before it cools down to its extremal state of Planck
mass\cite{Nicolini:2005vd,Isi:2013cxa}.  It was proposed this extremal
state as a candidate of dark matter\cite{Kovacik:2015yqa}.  In this
letter, we consider a different scenario that dark matter annihilation
in the gravitational field of noncommutative black holes.  At final
stage of evaporation, we hypothesize the existence of a thermal
equilibrium state composed of a burning black hole relics fueled by
dark matter accretion.
Note that it has been discussed that the accretion of primordial
black holes may be significant in the radiation-dominated era for
some types of modified theories of gravity and 
the extension of the lifetime is estimated \cite{PBH_Accretion}.
In contrast, we consider much late time equilibrium states of PBHs
in which the effect of noncommutativity might be significant.

In fact, the effect of noncommutativity in the early universe has
been considered in various contexts; for example, via the density fluctuation and the CMB power spectrum\cite{NC_CMB}.
This study, on the other hand, aims to shed light on a different aspects of the effect of
NG to cosmology; namely the late time effects through stabilized primordial black holes.
Since it is urged to find any kind of imprinting of Planck scale
physics in cosmology, it is worth considering another scenario here.

This paper is organized as follows: in the section
\ref{sec:fate-schw-black},
we review the
fate of noncommutative geometry inspired Schwarzschild black hole
(NCGS).  In the section \ref{sec:dark-matter-particle},
we calculate the dark matter distribution
in the vicinity of an extremal NCGS black hole.  In the section
\ref{sec:accr-vers-radi}, we
discuss the accretion model for a polytropic type of dark matter.  In
the section \ref{sec:anoth-equil-point},
we discuss the stable configuration of near-extremal
NCGS black hole.  
At last, we have comments and discussion in the
  section \ref{sec:discussion}.

\section{The fate of Schwarzschild black holes in noncommutative space}
\label{sec:fate-schw-black}

For a noncommutative space, one expects its coordinates do not commute and satisfy following relation:
\begin{equation}
[x^\mu,x^\nu] = i\theta \epsilon^{\mu\nu}.
\end{equation}
The formulation of coherent state in a complexified plane suggests a position measurement gives a smearing Gaussian distribution instead of the delta function\cite{Smailagic:2003rp}, that is
\begin{equation}
\delta(\vec{r}) \to \frac{1}{(4\pi \theta)^{3/2}}\exp \bigg( -\frac{r^2}{4\theta} \bigg).
\end{equation} 

Incorporate this smearing effect into mass distribution in the usual General Relativity, a noncommutative geometry inspired Schwarzschild (NCGS) black hole was constructed\cite{Nicolini:2005vd} and has the metric
\begin{eqnarray}
&&ds^2 = \bigg( 1-\frac{r_g(r)}{r} \bigg) c^2dt^2 
-\bigg( 1-\frac{r_g(r)}{r} \bigg)^{-1}dr^2-r^2(d\theta^2+\sin^2\theta d\phi^2),\nonumber\\
&&r_g(r) = \frac{4GM}{c^2\sqrt{\pi}}\gamma \bigg(\frac{3}{2},\frac{r^2}{4\theta} \bigg),
\end{eqnarray}
where the lower incomplete Gamma function is defined as
\begin{equation}
\gamma(s,x) \equiv \int_0^{x} dt \, t^{s-1} e^{-t}.
\end{equation}
It is straightforward to show that the Schwarzschild radius $r_g \to
\frac{2GM}{c^2}\equiv R_g$ is recovered as commutative limit, $\theta
\to 0$, is taken.  That the noncommutative black hole has two horizons
while its mass is larger than a critical value $M_c =
1.904\sqrt{\theta} c^2/G$ shows an interesting similarity to the
Reisner-Norstr\"{o}m black hole\cite{Kim:2008vi}.  The thermal
behavior of a NCGS black hole starts to gradually deviate from that of
a Schwarzschild black hole when the horizon $r_H\lesssim
6\sqrt{\theta}$.  The final stage of a Schwarzschild black hole could
have been violent and unpredictable according to Hawking's relation
$T_H = c^3/8\pi GM$.
 It is more likely that the Einstein's theory of general relativity
 would be replaced by a UV finite theory of quantum gravity toward the
 Planck scale.  Though a fully comprehensive theory of quantum gravity
 is still unavailable, its effect on spacetime might be captured by
 some effective theories.   With the naive help of GUP, the
 evaporation could be stopped at finite but still high
 temperature\cite{Adler:2001vs,Scardigli:2010gm}.   In contrast, the
 NCGS black holes cool down and reach its extremal state at $M=M_c$.
 These cold relics may not stay completely quiet since they could trap
 the surrounding abundant dark matter particles (DMP) and were ignited
 by the accretion.

\section{Dark Matter Particle Phase-Space Distribution}
\label{sec:dark-matter-particle}

In this section, we would like to derive the dark matter phase-space distribution in the vicinity of a extremal NCGS black hole.  The distribution around the Schwarzschild black hole has been discussed in  \cite{Baushev:2008yz} and here we generalized it to the case of extremal NCGS black hole.  We will assume those are non-interacting and non-relativistic DMPs and their speed are estimated a few hundred kilometers per second in the Galaxies.  The trajectory of a DMP of mass $m$ in the gravitational field of a NCGS black hole can be summarized in the following equations:
\begin{eqnarray}\label{trajectory}
&&ct = \frac{E_0}{mc^2}\int{\frac{dr}{(1-\frac{r_g}{r})\sqrt{(\frac{E_0}{mc^2})^2-(1-\frac{r_g}{r})(1+\frac{L^2}{m^2c^2r^2})}}},\nonumber\\
&&\phi = \int{\frac{Ldr}{r^2\sqrt{\frac{E_0^2}{c^2}-(m^2c^2+\frac{L^2}{r^2})(1-\frac{r_g}{r})}}},
\end{eqnarray}
where the conserved total energy $E_0\simeq mc^2$ and angular momentum
$L$ are define as
\begin{equation}
E_0 = \bigg( 1-\frac{r_g}{r} \bigg) mc^3 \dot{t},\qquad L = mr^2\dot{\phi}.
\end{equation}
The dot derivative is with respect to the proper time.  The radial speed component  $v_r$ and tangent one $v_t$ can be calculated from (\ref{trajectory}):
\begin{eqnarray}\label{velocity}
&&v_r = \sqrt{\frac{-g_{rr}}{g_{tt}}}\frac{dr}{dt} = \sqrt{\frac{r_g}{r}-\frac{\alpha^2}{r^2}(1-\frac{r_g}{r})},\nonumber\\
&&v_t = \sqrt{\frac{-g_{\phi\phi}}{g_{tt}}}\frac{d\phi}{dt}=\frac{\alpha}{r}\sqrt{1-\frac{r_g}{r}},
\end{eqnarray}
where we have adopted the natural unit $c=1$ and defined $\alpha \equiv \frac{L}{mc}$ for convenience, and DMP is also assumed to be non-relativistic.  Now we define the flux as number of particles crossing a sphere of fixed radius $r$ in unit proper time and unit solid angle, that is
\begin{equation}
dF = 4\pi r^2 {\cal N} v \cos{\theta} d\tau d\Omega,
\end{equation}
where ${\cal N}$ is the particle density.  After substituting 
\begin{equation}
\cos{\theta}d\Omega = \frac{\pi}{r_gr}(1-\frac{r_g}{r})d(\alpha^2), \qquad v=\sqrt{v_r^2+v_t^2}=
\sqrt{\frac{r_g}{r}},
\end{equation}
One obtains 
\begin{equation}
dF = 4\pi^2{\cal N}\frac{(r-r_g)^{3/2}}{\sqrt{r_g}r}d(\alpha^2)dt
\end{equation}
Assume our DMP detector locates at far distance $r_\infty$ where the space is asymptotic flat, this flux becomes
\begin{equation}
dF_\infty = \pi \frac{n_\infty}{v_\infty}d(\alpha^2)dt,
\end{equation}
for the number density per radius $n_\infty = \int{d\Omega} {\cal N}$ and DMP speed $v_\infty$ found at $r_\infty$.  To proceed, we will assume the flux per angular momentum and per time $\frac{dF}{d(\alpha^2)dt}$ remains constant at arbitrary distance.  This determines the DMP density distribution per solid angle
\begin{equation}
{\cal N} = \frac{n_\infty}{4\pi v_\infty}\frac{\sqrt{r_g}r}{(r-r_g)^{3/2}},
\end{equation}
for each given $n_\infty$ and $v_\infty$ found at our detector.  We remark that from (\ref{velocity}) only those DMPs with angular momentum $\alpha<2.408$ can reach the horizon and be caught by the extremal NCGS black hole. 

\begin{figure}[tbp]
\includegraphics[width=0.5\textwidth]{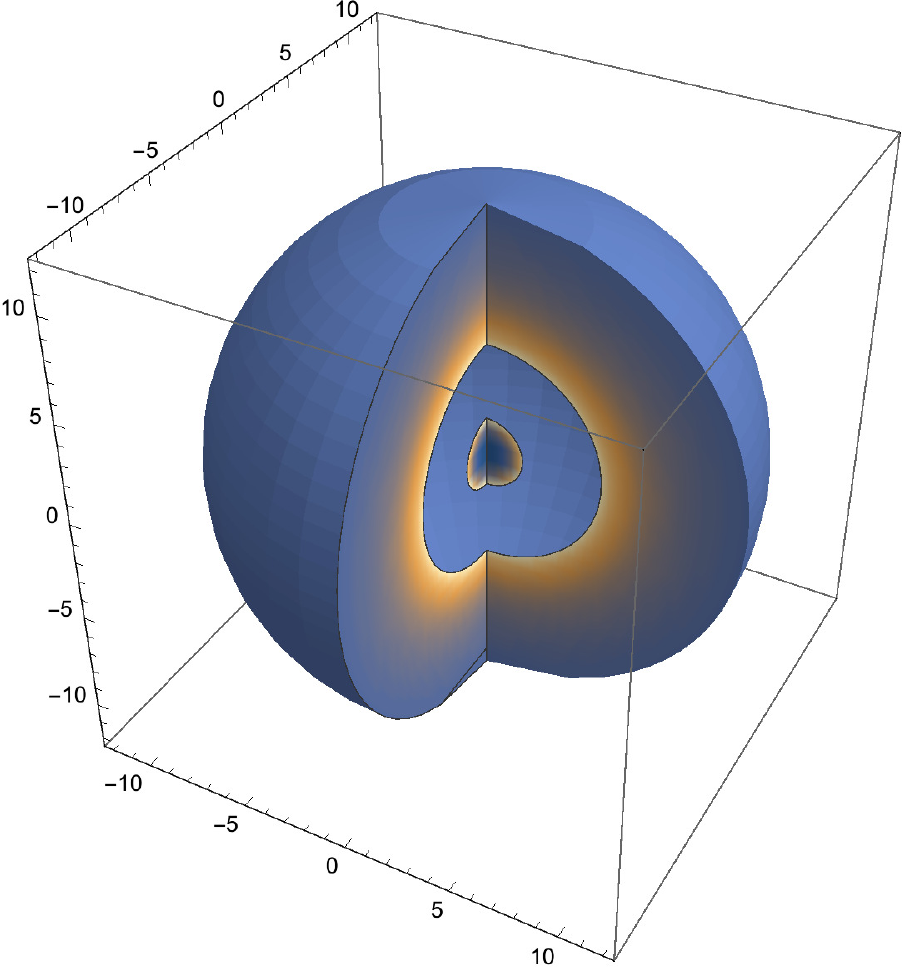}
\includegraphics[width=0.06\textwidth]{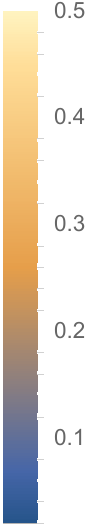}
\caption{\label{fig:density} Dark matter particle density distribution per solid angle versus the radial distance ($n_\infty$, $v_\infty$ are set to be 1 and $2GM/c^2$ is chosen as $4\sqrt{\theta}$.).
 We remark that while the distribution outside the (outer) horizon, located at $r_+=3.68\sqrt{\theta}$, is almost same as that of Schwarzschild black hole, there is nonzero distribution
inside the inner horizon, $r_-=2.48\sqrt{\theta}$, for the noncommutative black hole 
(but it may be just a formal distribution from the formula).
Particles are not allowed in the region between two horizons.}
\end{figure}

\section{Accretion versus Radiation}
\label{sec:accr-vers-radi}

Now one considers the model of dark matter accreted to the NCGS black hole.  The accretion model was derived in \cite{Michel:1972} and applied to many situations including the supermassive black holes\cite{Chakrabarti:1996cc,Peirani:2008bu}.  We assume the DMP behaves like an ideal fluid, i.e., $T_\mu^\nu = (P+{\cal E})u^\nu u_\mu -P\delta_\mu^\nu$.   The conservation of mass flux and energy-momentum flux give

\begin{eqnarray}
&&J^k{ }_{;k}=0 \longrightarrow Nv\sqrt{-g} = c_1;\\
&&T^k_{i;k}=0 \longrightarrow (P+{\cal E})g_{00} c \dot{t}v\sqrt{-g}=c_2.
\end{eqnarray}
Together we obtain
\begin{equation}
\bigg( \frac{P+{\cal E}}{N} \bigg) \bigg( 1-\frac{r_g}{r}+v^2 \bigg)^{1/2} = \frac{c_2}{c_1}.
\end{equation}
The differential relation between $v$ and $r$ is captured by the (solar) wind equation:
\begin{equation}
\frac{dv}{v}\big[ V^2F(r,v)-v^2 \big]
- \frac{dr}{2 r}\bigg[ \frac{r_g(r)}{r} -r_g'(r) -4V^2F(r,v) \bigg]=0,
\end{equation}
where
\begin{equation}\label{eqn:FV}
F(r,v)=1-\frac{r_g(r)}{r}+v^2, \qquad V^2 = \frac{d\log(P+{\cal E})}{d\log N}-1.
\end{equation}
The critical point flow ($v$ monotonically increases or decreases along the trajectory) occurs where both bracketed factors vanish simultaneously, namely
\begin{equation}
v_c^2 =
\frac{1}{4} \bigg(\frac{r_g(r_c)}{r_c}-r_g'(r_c) \bigg),
\qquad 
V^2\big|_{r_c} = \frac{v_c^2}{1-3 v_c^2 + r_g'(r_c)} .
\label{eq:crit_point}
\end{equation}
It is convenient to use the indicator of phase space density $Q$ for a self-similar radial infalling DMP\cite{Henriksen:2006gi}, where $Q \propto r^{-\beta}$.   A typical $\beta\simeq 1.87$ was found by \cite{Taylor:2001bq} for cluster-size halo.   The indicator is conserved during expansion of the universe and defined as
\begin{equation}
Q = \frac{mN}{\vsq^{3/2}}
\end{equation}
for velocity dispersion $\vsq$.  To be general, let us consider a polytropic gas of index $1/(\gamma-1)$ with total energy density and pressure:

\begin{equation}
{\cal E} = mNc^2 + \frac{P}{\gamma-1}, \qquad P = \frac{1}{3}mN\vsq c^2.
\end{equation}
One can further obtain the differential: 
\begin{equation}
\frac{d(P+{\cal E})}{dN} = mc^2 + \frac{5\gamma}{9(\gamma-1)}m c^2 
\bigg(\frac{N}{Q_N} \bigg)^{2/3}\,,
\end{equation}
for $Q_N\equiv Q/m$.  After Taylor expansion at large $Q_N$, equation (\ref{eqn:FV}) becomes
\begin{equation}
V^2 \simeq \frac{2\gamma}{9(\gamma-1)}\bigg(\frac{N}{Q_N} \bigg)^{2/3} 
-\frac{2\gamma^2}{27(\gamma-1)^2} \bigg(\frac{N}{Q_N} \bigg)^{4/3} 
+\cdots \,. 
\end{equation}
At far away from the black hole where $v_{\infty}\to 0$ for $r_{\infty} \gg r_g$, one obtains
\begin{equation}
\frac{c_2}{c_1}=mc^2 + \frac{\gamma}{3(\gamma-1)}m c^2 \bigg(\frac{N}{Q_N} \bigg)^{2/3}
\end{equation}
and\footnote{We notice a mistake in \cite{Peirani:2008bu} for the power in the ratio $N_c/N_\infty$ due to picking up a subleading term in the Taylor's expansion.} 
\begin{equation}
\bigg( \frac{N_c}{N_\infty} \bigg)^{4/3} \simeq \frac{2(\gamma-1)}{\gamma}\langle v_\infty^2 \rangle,
\qquad 
\langle v_c^2\rangle \simeq
\bigg( \frac{2(\gamma-1)}{\gamma}\langle v_\infty^2 \rangle\bigg)^{\frac{1}{2}}\,.
\label{eq:N_ratio}
\end{equation}
where we have assumed that $r_g'(r_c)$ is negligible. Since $r_g'(r)$ becomes exponentially small for $r > \sqrt{\theta}$, this is justified if $r_c$ is somehow large compared the location of degenerate horizon, $r \simeq 3.02\sqrt{\theta}$.
At last, $r_c$ is solved from the first relation of\eqref{eq:crit_point}.
Again, we assume that $r_g'(r_c)$ is negligible and $r_g(r_c)$ is taken as its asymptotic value
$r_g(\infty)=\frac{2GM}{c^2} \equiv R_g$, and the solution is found to be
\begin{equation}
r_c \simeq 
\frac{R_g}{4}\sqrt{\frac{\gamma}{2(\gamma-1) \langle v_\infty^2 \rangle}}\,.
\end{equation}
The accretion rate becomes
\begin{equation}
\frac{dM_{bh}}{dt} = 4\pi r_c^2 \frac{T_0^1}{c} =
\frac{\pi}{4} r_g(r)^2 c Q\,,
\end{equation}
where $\langle v_\infty^2 \rangle \ll 1$ and $r_g'(r_c) \ll 1$ are assumed.
Taking account of Hawking radiation
\begin{equation}
\frac{dM_{bh}}{dt} =  
\frac{\pi}{4} R_g^2 c Q
- c^{-2}(4\pi r_H^2)\epsilon \sigma T^4, 
\end{equation}
where the Stefan-Boltzmann constant $\sigma = \frac{\pi^2
  k_B^4}{60\hbar^3 c^2}$ and assuming $\epsilon=1$ for perfect
blackbody.  $r_H$ is the location of the outer horizon, a solution of
$r_g(r_H)=r_H$.  To the leading order calculation, we can approximate $r_H \simeq R_g$.  A careful treatment with subleading correction is given in the next section.
   This implies a unique temperature in the thermal equilibrium
\begin{equation}\label{remnant_temp}
T_{rem} = 
\frac{1}{2} \bigg(\frac{c^3 Q}{\sigma} \bigg)^{\frac{1}{4}} \,.
\end{equation} 
We have following remarks: at first, the fact that equilibrium temperature (\ref{remnant_temp}) is independent of $\theta$ implies that we simply obtained the solution of Schwarzschild counterpart.  Namely the difference between a noncommutative black hole and an ordinary black hole is hardly observed at this equilibrium point. Secondly, this NCGS black hole immersed in the dark matter halo with typical phase-space density $Q \approx 3.51\times 10^{-9} M_{\odot}pc^{-3}km^{-3}s^{3}$  would burn at a temperature\footnote{%
The phase-space density of halo is well fitted by the relation\cite{Peirani:2007cj}
\begin{equation}
Q \approx \frac{3.51\times 10^{-9}}{M_{11}^{1.54}}M_{\odot}pc^{-3}km^{-3}s^{3},
\end{equation}
for halo mass $M_{11}$ in the units of $10^{11}M_{\odot}$.
It should also be noted that 
since we have used a dimensionless $\langle v^2 \rangle$ to derive \eqref{remnant_temp},
we need to put an extra $c^3$ in the parenthesis of the formula.}
$T_{rem} \approx 1.17 \times 10^5 \, \mathrm{K}$, which could look
like  a hot spot in the present cosmic microwave background but
difficult to be identified due to its small size, about $10^{-9}$m.\footnote{%
Now we can use the formula for conventional black
  holes.
The equilibrium temperature is
\begin{align}
  T_H= \frac{T_P}{4\pi} \frac{\ell_P}{R_g} = 1.17 \times 10^{5} K \,.
\end{align}
By use of $T_P=1.42 \times 10^{32}$K and $\ell_P=1.62 \times
10^{-35}$m,
the size of the black hole is $R_g=9.64\times 10^{25} \ell_P\simeq
1.56 \times 10^{-9}$m. } This is huge compared to Planck size, but
microscopic in the context of cosmology.
The total mass of the black hole is $M = 4.82 \times 10^{25}
m_P=1.05\times 10^{18}$kg, or equivalent $10^{-12} M_\odot$.
This black hole has longer life time than the age of universe, and in the
range of proposed primordial black hole as dark matter candidate.
At last, a conventional Schwarzschild may also reach a thermal equilibrium with surrounding dark matter at the temperature predicted in (\ref{remnant_temp}), when the radiation and accretion rate are the same.  However, this equilibrium is unstable for the following reason:  while the mass of a black hole decreases, the accretion rate also
decreases.  As a result the Hawking temperature increases for its negative specific heat.
In contrast, the temperature of NCGS black hole may behave like a conventional black hole at large mass, but it starts to decrease after reaching the
maximum temperature $T_\text{max}\simeq 0.015/\sqrt{\theta}$ and then
cool down as a remnant.  We then expect to have an additional stable
equilibrium point for NCGS black hole at a much smaller mass. The
detailed analysis will be carried out in the section \ref{sec:anoth-equil-point},
and there it will be found that the other equilibrium takes place at $T\simeq 10^7$K, very close to its extremal state.
Therefore at this equilibrium point, the Planck-size NCGS black hole has a much cooler temperature than its Schwarzschild companion.  The total
flux is also much smaller compared the previous equilibrium case.
From the discussion above, this equilibrium is expected to be stable against evaporation.

\section{Another equilibrium point near the extremal one}
\label{sec:anoth-equil-point}

In the previous section, we have considered the equilibrium configuration of
PBH and dark matter halo through Hawking radiation and accretion.
In this section, we look at an equilibrium point which is very close
to the extremal point of NCBH.
We will keep $r'(r_c)$ and also the difference between $r_H$ and
$R_g$ (the gravitational radius for the total mass), which are
neglected in the previous analysis.

With keeping $r'(r_c)$ in deriving \eqref{eq:N_ratio}, we find a correction,
\begin{align}
  \bigg( \frac{v_c}{v_\infty} \bigg)^2
\simeq &
\frac{2(\gamma-1)}{\gamma} \frac{1}{\sqrt{1+r_g'(r_c)}} \frac{1}{\vev{v_\infty^2}}
+ \frac{6(\gamma-1)^2}{\gamma^2} \bigg(\frac{1}{\sqrt{1+r_g'(r_c)}}-1 \bigg)
\frac{1}{\vev{v_\infty^2}^2} \,.
\end{align}
$r_c$ is defined as a solution of the following relation, 
\begin{align}
  r_c = &\frac{R_g}{2(4v_c^2+r_g'(r_c))} 
\left[
1+\sqrt{1- \frac{8\theta r_g'(r_c)}{R_g^2} \left( 4v_c^2+r_g'(r_c) \right)}
\right]
\nn\\\simeq &
\frac{R_g}{4v_c^2+r_g'(r_c)} 
\left[
1- \frac{2\theta r_g'(r_c)}{R_g^2} \left( 4v_c^2+r_g'(r_c) \right)
\right] \,,
\end{align}
where we have taken the first order in $\theta$.
When $r_c$ is sufficiently large compared to $2\sqrt{\theta}$, we may
take $r_g'(r_c) \leq {\cal O}(\theta)$, and
\begin{align}
  r_c \simeq & \frac{R_g}{4v_c^2} \bigg[ 1+ \frac{r_g'(r_c^{(0)})}{4v_c^2} \bigg] 
\end{align}
where $r_c^{(0)}=R_g/4v_c^2$ is $\theta$ independent leading order solution.
Thus, with keeping the correction terms, the accretion rate is given
by
\begin{align}
  \frac{dM_{bh}}{dt} =&
\frac{\pi}{4}R_g^2 c Q \cdot
\bigg(1+ \frac{r_g'(r_c^{(0)})}{4v_c^2}\bigg)^{2}
\bigg(1+ \frac{1}{3}\frac{\gamma}{\gamma-1} \vev{v_\infty^2} \bigg) \,.
\end{align}

Now we move on to the correction to Hawking radiation.
The location of the horizon, a solution of $r_g(r_H)=r_H$, is
approximately
\begin{align}
  r_H\simeq  R_g - \frac{2\theta}{R_g} r_g'(R_g) \,.
\end{align}
Hawking temperature $T_H$ is given by \cite{Nicolini:2005vd}
\begin{align}
  T_H=T_P \ell_P \bigg[ \frac{1}{4\pi} \frac{dg_{00}}{dr} \bigg]_{r=r_H}
=\frac{\hbar c}{4\pi k_B r_H} \bigg[1-\frac{R_g}{2\sqrt{\pi}}\frac{r_H^2}{\theta^{3/2}}e^{-\frac{r_H^2}{4\theta}} \bigg]
\,.
\end{align}
Then the condition for equilibrium is 
\begin{align}
  \frac{dM_{bh}}{dt} =&
\frac{\pi}{4}R_g^2 c Q \cdot
\bigg(1+ \frac{r_g'(r_c^{(0)})}{4v_c^2}\bigg)^{2}
\bigg(1+ \frac{1}{3}\frac{\gamma}{\gamma-1} \vev{v_\infty^2} \bigg)
- \frac{4\pi r_H^2}{c^2} \epsilon \sigma T_H^4
\nn\\=& 0 \,,
\label{eq:eq_cond}
\end{align}
and the solution is
\begin{align}
  T_H =& \frac{1}{2} \bigg( \frac{c^3 Q}{\sigma}\bigg)^{1/4}
\sqrt{\frac{R_g}{r_H}} \bigg(1+ \frac{r_g'(r_c^{(0)})}{4v_c^2}\bigg)^{1/2}
\bigg(1+ \frac{1}{3}\frac{\gamma}{\gamma-1} \vev{v_\infty^2} \bigg)^{1/4}
\,.
\end{align}

The extremal NCBH is realized if the mass coincides with the extremal
mass, $M_\text{ext}=1.904 m_P\sqrt{\theta \ell_P^{-2}}$ where
$m_P$ and $\ell_P$ are Planck mass and length respectively, and
the location of the degenerate horizon is $r_0=3.02\sqrt{\theta}$.
We now examine the case in which the total mass of the NCBH is very
close to $M_\text{ext}$.
By setting $r_H=r_0 + \epsilon$, the Hawking temperature is
$T_H=T_P \tilde{\epsilon}\tilde{\theta}^{-1/2} \big( 0.0224 - 0.00744
\tilde{\epsilon} + {\cal O}(\tilde{\epsilon}^2) \big)$
where we have introduced dimensionless $\tilde{\epsilon}$ and
$\tilde{\theta}$ as $\tilde{\epsilon}=\epsilon/\sqrt{\theta}$
and $\tilde\theta=\theta\ell_P^{-2}$, and $T_P$ is Planck temperature.
As seen from this expression, the extremal limit $r_H \rightarrow r_0$
($\epsilon\rightarrow 0$)
and the commutative limit $\theta\rightarrow 0$ are not commuting.
By using the leading order part of $T_H$ and also dropping the
correction term for the accretion rate, the equilibrium condition is
\begin{align}
  \frac{\pi c}{4} R_g^2 Q=& \frac{4\pi \sigma}{c^2} r_0^2 \frac{\tilde{\epsilon}^4 \hbar^4 c^4}{\theta^2 k_B^4}
 \times (0.0224)^4 
 \,,
\end{align}
If we adopt the same typical $Q$ value in the main part, we find
\begin{align}
  M = 1.111 \times 10^{51}
  \frac{\tilde{\epsilon}^2}{\sqrt{\tilde{\theta}}}m_P \,.
\end{align}
This value has to be very close to the extremal mass $M_\text{ext}$
under the assumption we take, which implies
$\tilde{\epsilon}\simeq 4.14 \times 10^{-25} \sqrt{\tilde{\theta}}$.
If we choose $\tilde{\epsilon}$ to this value, the Hawking temperature
is $T_H \simeq 0.927 \times 10^{-25} T_P=1.31 \times 10^7$K
and the total mass of NCBH is $M \simeq M_\text{ext}$.

We remark that a thermally stable black hole relics cannot exist without the noncommutativity.  Otherwise, accretion would be impossible due to the furious circumference heated up by a usual Schwarzschild black hole of Planck size.

\section{Discussion}
\label{sec:discussion}

So far, our simple model only discussed the equilibrium between accretion of dark matter
halo and the Hawking radiation. 
For the stable equilibrium, the mass of black hole is almost Planck mass
$10^{-8}$kg
and its size is of almost Planck size,
while for the unstable equilibrium (namely for the conventional
Schwarzschild black hole), the mass is about $10^{18}$kg
($10^{-6}$ of Earth mass or $10^{-4}$ of the mass of the Moon)
and the size is of order $10^{-9}$m.
Though the gravitational field near the horizon is not so small
(about $10^{13}$ stronger compared to typical neutron stars for the unstable
case, $10^{40}$ for the stable case), their total gravitational
field and the flux are quite small.
For the stable configuration, it seems to not form a dark matter halo
around them and the equilibrium may be realized when they happen to
locate in a region of high energy density.
As for the unstable case, the situation is subtle and
it would be necessary to reexamine the applicability of the same formula of accretion
rate.

A lower bound of $\theta$ could be estimated by noticing that the
highest temperature which can be reached by a noncommutative black
hole is given by
$T_{max}=\frac{0.015}{\sqrt{\theta}}$.  Further
asking the wavelength of Hawking radiation should not be shorter than
the Planck length, one could estimate the lower bound $\theta \ge 10 \l_p$\cite{Nicolini:2005vd}.

The production rate for primordial black holes are considered in \cite{Carr:2005bd}, but they mainly focused on the ones heavier than $10^{15}$g, as light ones have already evaporated in the current universe.
Since the noncommutativity makes black holes stable, and then if the production rate for light primordial black holes are sufficiently high, we would have a upper limit for possible $\theta$.
The mass of the extremal black hole is given by $1.9 \sqrt{\theta}/G$ which is of order Planck mass ($10^{-8}$g).

Recent study has suggested possible connection between the Cosmic Infrared Background (CIB) and primordial black hole \cite{Kashlinsky:2016sdv}.  Those near-IR fluctuation of wavelength $2 - 5\mu m$ could have been majorly contributed from objects with  temperature $580-1450$K according to Wien's displacement law, which could be ordinary primordial black holes with mass among $4 -11\times 10^{-11} M_{\odot}$.  
If some of those unidentified sources were from extremal black hole
remnants, they should have larger number density but much lighter
individual mass. 
For instance, the radiation flux of an ordinary primordial black hole
of mass $M_h$ $\propto (2GM_h)^2 (8\pi G M_h)^{-4}$; but if same
amount mass were made of extremal NCGS black hole at same temperature,
the flux $\propto (M_h/M_{ext})(2GM_{ext})^2(8\pi G M_h)^{-4}$.  The
latter is much weaker than the former by a factor $M_{ext}/M_h \sim
10^{-26}$.
The abundance of these extremal remnant might modify our resolutions toward some puzzles in the modern cosmology.  Radiation come from those warm remnants may accelerate local reionization of hydrogens after the Big Bang to clear the {\sl foggy} universe.  Those remnants may also help early galaxies formation by serving as seeds.  Inspection of the abovementioned phenomenological application is under progress and will be reported in a separated paper.   

\begin{acknowledgments}
We are grateful to Pei-Ming Ho and Je-An Gu for his helpful discussion.  This work is supported in parts by the Chung Yuan Christian University, the Taiwan's Ministry of Science and Technology (grant No. 102-2112-M-033-003-MY4) and the National Center for Theoretical Science. 
\end{acknowledgments}



\end{document}